\documentclass[aps,floatfix,showpacs,epsfig]{revtex4}
\usepackage{epsfig}
\usepackage{natbib}
\usepackage{graphicx}
\usepackage{amssymb}
\usepackage{amssymb,latexsym}
\usepackage[mathscr]{eucal}
\def\beq{\begin{equation}}
\def\eeq{\end{equation}}
\def\ba{\begin{eqnarray}}
\def\ea{\end{eqnarray}}
\begin{document}
\title{\large \bf Radiation from the LTB black hole }
\author{J. T. Firouzjaee}
\affiliation{Department of Physics, Sharif University of Technology,
Tehran, Iran} \email{firouzjaee@physics.sharif.edu; taghizade.javad@gmail.com}

\author{Reza Mansouri}
\affiliation{Department of Physics, Sharif University of Technology,
Tehran, Iran and \\
  School of Astronomy, Institute for Research in Fundamental Sciences (IPM), Tehran, Iran}
 \email{mansouri@ipm.ir}
\begin{abstract}
Does a dynamical black hole embedded in a cosmological FRW background emit the Hawking radiation where
a globally defined event horizon does not exist? What are the differences to the Schwarzschild black hole?
What about the first law of black hole mechanics? We face these questions using the LTB cosmological
black hole model recently published. Using the Hamilton-Jacobi- and radial null geodesic- methods suitable
for dynamical cases, we show that it is the apparent horizon which contributes to the
Hawking radiation and not the event horizon. The Hawking temperature is calculated using the two different methods
giving the same result. The first law of LTB black hole dynamics and the thermal
character of the radiation is also dealt with.

\end{abstract}
\maketitle

\section{Introduction}

Any black hole in the real universe is necessarily a dynamical one, i.e. it
is neither stationary nor asymptotically flat. Therefore, its horizon has to be defined locally. The
need to understand such dynamical and cosmological black holes
has led to a revival of discussions on the concepts of black hole itself, its
singularity, horizon, and thermodynamics {\cite{haw-bh}. Indeed, the
conventional definition of black holes implies an asymptotically flat
space-time and a global definition of the event horizon. The universe, however, is not
asymptotically flat and a global definition of the horizon is
not possible. The need for local definition of black holes and their horizons
has led to concepts such as Hayward's trapping horizon {\cite{Hayward94}, and
Ashtekar and Krishnan's dynamical horizon {\cite{ashtekar02}. It is not a trivial fact that in
a specific example of a dynamical black hole any of these horizons may occur. In addition, we
do not know if and how the Hawking radiation and the laws of black hole thermodynamics will apply
to dynamical and cosmological black holes. Hence, it is of special interest to find out specific
examples of dynamical black holes as a test bed for horizon problems. \\

Within a research program, we have already found an exact solution of Einstein equations based
on the LTB solution {\cite{LTB} representing a dynamical black hole which is asymptotically
FRW {\cite{man}. Different horizons and local mass definitions applied to this cosmological
black hole are also reported {\cite{man-g,man-ash}. In this paper we are looking into the question
of the Hawking radiation \cite{Hawking2} and the first law of black hole dynamics in such a
dynamical LTB black hole .\\

A first attempt to look into Hawking radiation from a 'cosmological black hole' is reported
in \cite{saida}. There the authors consider the Einstein-Straus solution and the Sultana-Dyer one as
cosmological black holes. The Einstein-Straus solution, however, is constructed such that it
represents a 'freezed-out' Schwarzschild black hole within a FRW universe; it can not represent
a dynamical black hole. The Sultana-Dyer solution is reproduced by the Schwarzschild metric through
a conformal factor; it represent a FRW universe with a fixed black hole of a certain mass within
it. Again, it can not reflect the features we expect from a dynamical black hole with the mass in-fall
within an asymptotically FRW universe. Therefore, it is still an open question how the Hawking
radiation and black hole dynamical laws will look like for a dynamical black hole within a FRW universe.
We will look into this question in the case of the LTB cosmological black hole we have found as an
exact solution of Einstein equations with mass in-fall, and without using a cut-and-paste technology
of manifolds {\cite{man}. \\
Now, Hawking's approach \cite{Hawking2} based on quantum field theory, and applied to
quasi cosmological black holes\cite{saida} is not a suitable method to calculate Hawking
temperature in the case of proper dynamical black holes where one has to solve the field equations
in a dynamical background. In such cases, like the LTB black hole {\cite{man}, one should look for alternative
approaches allowing to calculate the temperature of the Hawking radiation and the surface gravity. The so-called
Hamilton-Jacobi (H-J) approach \cite{svmp} has initiated methods suitable in such cases. The
method, however, suffers from not being manifestly covariant. Using Kodama's formalism \cite{Kodama:1979vn}
for spherically symmetric space-times, and based on the Hamilton-Jacobi method, Hayward \cite{Hayward:1997jp}
has formulated a covariant form of the H-J approach and studied the quantum instability of a
dynamical black hole \cite{Hayward:2008jq,hay0906}.\\

Another useful method is the semi-classical tunneling  approach to Hawking radiation due to
Parikh and Wilczek (PW) \cite{parikh,P,svmp} using radial null geodesics. We will apply this
radial null geodesic method, formulated originally for the static case, to our dynamic
cosmological black hole and compare the result with that of the Hamilton-Jacobi method. The method, written
in a gauge-invariant form, has led first to a Hawking temperature twice as large as the correct
one \cite{Chowdhury-akh06}. It has, however, been shown that the correct Hawking temperature is regained by taking into account
a contribution from the time coordinate upon crossing the horizon \cite{akh08-pilling}. \\

Among different definitions for the surface gravity of evolving horizons proposed in the past, the one
formulated by Hayward \cite{Hayward:1997jp} based on the Kodama's formalism \cite{Kodama:1979vn}
is the most suitable one to be used in our dynamic case. This leads us to the first law of black holes compatible
with the Hawking temperature calculated by two different methods mentioned above.\\
The question of thermal character of the Hawking radiation
will also be discussed within this formalism. As Parikh and Wilczek have already pointed
out, the Hawking radiation is non-thermal when the energy conservation
is enforced \cite{kraus95,parikh}. As a result it has been shown by Zhang et al
\cite{cai-inf09} that the total entropy becomes conserved and the black hole evaporation process is
unitary.\\
We will introduce in section II the LTB black hole. In Section III, the covariant Hamilton-Jacobi tunneling
method is introduced and applied to the LTB black hole. Section IV is devoted to
the radial null geodesic method and its application to the LTB black hole. In Section V, the
first law for the LTB black hole is derived. The thermal character of radiation is considered in
section VI. We then conclude in section VII.

 \section{Introducing the LTB black hole}

The cosmological LTB black hole is defined by a cosmological
spherical symmetric isotropic solution having an overdense mass
distribution within it {\cite{man}. The overdense mass distribution collapses due to the
dynamics of the model leading to a black hole at the center of
 the structure.  The LTB metric is the simplest
 spherically symmetric solution of Einstein equations
representing an inhomogeneous dust distribution {\cite{LTB}. It may be written in synchronous coordinates as
\begin{eqnarray}\label{ltbe000}
 ds^{2}= - dt^{2}+\frac{R'^{2}}{1+f(r)}dr^{2}+R(t,r)^{2}d\Omega^{2},
\end{eqnarray}
representing a pressure-less perfect fluid satisfying
\begin{eqnarray}\label{ltbe00}
\rho(r,t)=\frac{2M'(r)}{ R^{2}
R'},\hspace{.4cm}\dot{R}^{2}=f+\frac{2M}{R}.
\end{eqnarray}
Here dot and prime denote partial derivatives with respect to the
parameters $t$ and $r$ respectively. The angular distance $R$,
depending on the value of $f$, is given by
\begin{eqnarray}\label{ltbe1}
R=-\frac{M}{f}(1-\cos \eta(r,t)),\nonumber\\
\hspace{.8cm}\eta-\sin \eta=\frac{(-f)^{3/2}}{M}(t-t_{b}(r)),
\end{eqnarray}
\begin{eqnarray}\label{dotltbe1}
\dot{R}=(-f)^{1/2}\frac{sin(\eta)}{1-cos\eta},
\end{eqnarray}
for $f < 0$,
 and
\begin{equation}\label{ltbe2}
R=(\frac{9}{2}M)^{\frac{1}{3}}(t-t_{b})^{\frac{2}{3}},
\end{equation}
 for $f = 0$, and
\begin{eqnarray} \label{ltbe3}
R=\frac{M}{f}(\cosh \eta(r,t)-1),\nonumber\\
\hspace{.8cm}\sinh \eta-\eta=\frac{f^{3/2}}{M}(t-t_{b}(r)),
\end{eqnarray}
for $f > 0$.\\
The metric is covariant under the rescaling
$r\rightarrow\tilde{r}(r)$. Therefore, one can fix one of the three
free functions of the metric, i.e. $t_{b}(r)$, $f(r)$, or $M(r)$.
The function $M(r)$ corresponds to the Misner-Sharp mass in general
relativity \cite{man-g}. The $r$ dependence of the
bang time $t_{b} (r)$ corresponds to a non-simultaneous big bang- or
big-crunch-singularity. \\
There are two generic singularities of this metric, where the
Kretschmann scaler and Ricci one become infinite: the shell focusing
singularity at $R(t,r)=0$, and the shell crossing one at
$R'(t,r)=0$. However, there may occur that in the case of $R(t,r)=0$
the density $\rho = \frac{M'}{R^{2}R'}$ and the term $\frac{M}{R^3}$
remain finite. In this case the Kretschmann scalar remains finite
and there is no shell focusing singularity. Similarly, in the
case of vanishing $R'$ the term $\frac{M'}{R'}$ may remain finite leading to a
finite density and no shell crossing singularity either.  Note that an expanding universe means generally $\dot{R}>0$.
However, in a region around the center it may happen that
$\dot{R}<0$, corresponding to the collapsing region.\\
The LTB metric may also be written in a form similar to the
Painlev$\acute{e}$ form of the Schwarzschild metric. By taking the
physical radius as a new coordinate using the relation
$dR=R'dr+\dot{R}dt$ one obtains
 \ba
 ds^{2}= (\frac{\dot{R}^{2}}{1+f}-1)dt^{2}+\frac{dR^{2}}{1+f}-\frac{2\dot{R}}{1+f}dR dt\nonumber\\+R(t,r)^{2}d\Omega^{2}.
\label{ltbph} \ea
The $(t,R)$ coordinates are usually called the physical coordinates. In the
case of $f = 0$, the metric is quite similar to the Painlev$\acute{e}$ metric.\\

\section{Hamilton-Jacobi method}\label{tunneling}

The Hamilton-Jacobi method to calculate the Hawking radiation uses the fact that
within the WKB approximation the tunneling probability for the
classically forbidden trajectory from inside to outside the
horizon is given by
\begin{equation}
\Gamma \propto \exp\left(- \frac{2}{\hbar}\mbox{Im } S\right),
\label{prob}
\end{equation}
where $S$ is the classical action of the (massless) particle to the
leading order in $\hbar$ \cite{svmp}. Note that the exponent has to
be a scalar invariant, otherwise no physical meaning could be given to
$\Gamma$. If, in particular, it has the form of a thermal emission
spectrum with $2 \mbox{Im }\, S=\beta \omega $, then both the
inverse temperature $\beta$ and the particle's energy $\omega$ have
to be scalars; otherwise no invariant meaning could
be given to the horizon temperature, which would then not be an observable.\\

Now, let us use the Kodama formalism to introduce the necessary
invariant quantities\cite{Hayward:2008jq}. Any spherically symmetric
metric can be expressed in the form \beq \label{metric} ds^2
=\gamma_{ij}(x^i)dx^idx^j+ R^2(x^i) d\Omega^2\,,\qquad i,j \in
\{0,1\}\;, \eeq
 where the two-dimensional metric
 \beq
d\gamma^2=\gamma_{ij}(x^i)dx^idx^j \label{nm} \eeq is referred to as
the normal metric, $x^i$ are associated coordinates, and $R(x^i)$ is
the areal radius considered as a scalar field in the normal
two-dimensional space. Another relevant scalar quantity on this
normal space is \beq \chi(x)=\gamma^{ij}(x)\partial_i R\partial_j
R\,. \label{sh} \eeq The dynamical trapping horizon, $H$, may be
defined by \beq \chi(x)\Big\vert_H = 0\,, \qquad \partial_i\chi
\Big\vert_H \neq 0\,. \label{ho} \eeq The Misner-Sharp gravitational
mass is then given by
 \beq M(x)=\frac{1}{2} R(x)\left(1-\chi(x)
\right)\, , \label{MS} \eeq
which is an invariant quantity on the
normal space. In the special case of the LTB metric this reduces to the
Misner-Sharp mass $M(r)$.  Note also that on the horizon $M\vert_H= m = R_H/2$.
Now, there is always possible is such a spherically symmetric space time to define 
a preferred observer and its related invariant energy corresponding to the 
classical action of a particle. The Kodama vector \cite{Kodama:1979vn}, representing 
a preferred observer corresponding to the Killing vector in the static case, for 
the case of the LTB metric (\ref{metric})is given by
\beq K^i(x)=\frac{1}{ \sqrt{-\gamma}}\varepsilon^{ij}\partial_j R\,,
\qquad K^\theta=0=K^\varphi \label{kodama} \;. \eeq
 Using this Kodama vector, we may introduce the invariant energy 
associated with a particle by means of its classical action being a 
scalar quantity on the normal 
space \beq \label{e} \omega = K^{i}\partial_i S . \eeq
Note that during the process of horizon tunneling $\omega$ is
invariant independent of coordinates and is regular across the horizon.
 In the case of the eikonal approximation for massless 
wave field (geometric optics limits), which plays an important role in calculating 
the Hawking radiation using tunneling method \cite{man-nielsen}, the classical action $S$  for the
massless particle satisfies the Hamilton-Jacobi equation \beq
\label{hj} \gamma^{ij}\partial_i S
\partial_j S = 0\,. \eeq
\\
The relevant imaginary part of the classical action along the $\gamma$ {\it null curve}
is calculated in \cite{Hayward:2008jq}, where it has
been shown that the tunneling rate (\ref{prob}) is valid for the future
trapped horizon, and \beq \mbox{Im }\, S = \mbox{Im }\,
\left( \int_{ \gamma} dx^{i}
\partial_i S \right)=\frac{\pi\omega_H}{\kappa_H}\,, \label{ima}
\eeq
where  $\omega_H$ is the Kodama energy and $\kappa_H$ is 
the dynamical surface gravity associated with the dynamical horizon:
\beq \kappa_H=\frac{1}{2}\Box_{\gamma} R
\Big\vert_H=\frac{1}{2\sqrt{-\gamma}}\partial_i(\sqrt{-\gamma}
\gamma^{ij}\partial_jR)\Big\vert_H. \label{surfaceg}
\eeq 
 These are scalar quantities in the normal space.
Therefore, the leading term of the tunneling rate is invariant, as it should be
for an observable quantity. The particle production rate then takes
the thermal form $ \Gamma \sim e^{-\frac{w}{T}}$ with
\beq T=\frac{\hbar \kappa_H}{2\pi}. \eeq \\

\subsection{Application to the LTB back hole}

Assume the cosmological LTB black hole which has an infinite redsift surface satisfying 
the eikonal approximation condition for the Hamilton-Jacobi equation (\ref{hj}). 
It has been shown in \cite{man-ash} that the apparent horizon of this LTB black hole 
in its last stages is a slowly evolving horizon with the least mass in-fall due to the 
expanding background preventing the mass infall to the central black hole. \\
Now, from equations (\ref{ltbe000}) and (\ref{ltbe00}), and the definition of the
surface gravity (\ref{surfaceg}), we obtain

\beq
\kappa_H=\frac{m}{R^2}-\frac{m'}{2RR'}=\frac{1}{2R}-\frac{m'}{2RR'},\label{sgrav}
\eeq
where $m$ is the Misner-Sharp mass on the horizon.  Using this expression for the surface gravity, we obtain
the Hawking temperature according to the Hamilton-Jacobi tunneling
approach:
\begin{eqnarray}
T&=&\frac{\hbar
\kappa_H}{2\pi}=\frac{\hbar}{4\pi}\frac{\sqrt{1+f}}{R'}
[-\frac{\partial_{t}(\dot{R}R')}{\sqrt{1+f}}+\partial_{r}(\sqrt{1+f})]\nonumber\\&=&
\frac{\hbar}{4\pi}(\frac{f'}{2R'}-\ddot{R}-\frac{\dot{R}\dot{R}'}{R'})
= \frac{\hbar}{4\pi}(\frac{1}{R}-\frac{m'}{RR'}). \label{ltb-t}
\end{eqnarray}
To relate this result to the temperature seen by the Kodama observer, we
calculate first the Kodama vector (\ref{kodama}). It is given by
$K^{i}=\frac{\sqrt{1+f}}{R'}(R',-\dot{R})$. The equation
$|K|=\sqrt{-K_{i}K^{i}}=\sqrt{1+f-\dot{R}^2}=\sqrt{1-\frac{2m}{R}}$
shows that the Kodama vector is a null vector on the horizon. The
corresponding velocity vector is then given by
$\hat{K}^{i}=\frac{K^{i}}{|K|}$. We then obtain the frequency
measured by such an observer as $\hat{\omega} = \hat{K}^{i}\partial_i S $. The emission rate will take the thermal form
\begin{equation}
\Gamma \propto e^{-\frac{\hat{w}}{\hat{T}}}
\end{equation}
which defines the temperature $\hat{T}$.
It is then easily seen that the
temperature for this observer is  \beq
\hat{T}=\frac{T}{\sqrt{1-\frac{2m}{R}}}, \eeq
which diverges at the horizon. The invariant redshift factor $\frac{1}{\sqrt{1-\frac{2m}{R}}}$ is 
the same factor which appear in the light frequency on the horizon showing an 
infinite redshift to the observer in the infinity. Then $T$ itself, being finite at the horizon, may be
interpreted as the redshift-renormalized temperature \cite{Hayward:2008jq}.

\section{ Radial null geodesic approach}

There is another approach to the Hawking radiation as a quantum
tunneling process using WKB approximation for radial null geodesics
tunneling out from near the horizon \cite{parikh}. The imaginary
part of the action is defined by
\begin{eqnarray}
{\textrm{Im}} S&=&
{\textrm{Im}}\int_{r_{\textrm{in}}}^{r_{\textrm{out}}}p_r
dr={\textrm{Im}}\int_{r_{\textrm{in}}}^{r_{\textrm{out}}}\int_0^{p_r}
dp'_r dr \nonumber
\\
&=&{\textrm{Im}}\int_{r_{\textrm{in}}}^{r_{\textrm{out}}}\int_0^H\frac{-dH'}{\dot
r} dr, \label{1.014}
\end{eqnarray}
using the Hamilton equation $\dot r=\frac{dH}{dp_r}|_r$ with $H$
being the Hamiltonian of the particle, i.e. the generator of the
cosmic time t. Now, taking the tunneling probability as $\Gamma\sim
e^{-\frac{2}{\hbar}{\textrm{Im}}S}$, being proportional to the
Boltzmann factor $e^{-\frac{\omega}{T}}$, we find the Hawking
temperature as
\begin{eqnarray}
T_H=\frac{\omega\hbar}{2{\textrm{Im}}S}. \label{1.021}
\end{eqnarray}
It is easy to show that for a Schwarzschild black hole one obtain
the correct expression $T_H=\frac{\hbar}{8\pi M}$. \\

It has been pointed out in \cite{Chowdhury-akh06}, that
$2Im\int_{r_{in}}^{r_{out}} p_rdr$ is not canonically invariant and
thus it does not represent a proper observable. The object which is
canonically invariant is $Im\oint p_rdr$, where the closed path goes
across the horizon and back. Using this invariant definition and
$\Gamma\sim e^{\frac{-1}{\hbar}Im\oint p_r dr}$, the Hawking
temperature is found to be twice the original temperature. This
discrepancy in the temperature has been resolved by considering
a temporal contribution to the tunneling amplitude. In the case of the
Schwarzschild black hole, the temporal contribution to the action is
found by changing Schwarzschild coordinates into Kruskal-Szekeres
coordinates and then matching different Schwarzschild time
coordinates across the horizon \cite{akh08-pilling}.

\subsection{Application to the  LTB black hole}

We use the LTB space-time in physical coordinates. In this case the outgoing
and ingoing null geodesics are given by
\begin{equation}
\frac{dR}{dt}=(\dot{R} \pm \sqrt{1+f}),
\end{equation}
where the plus sign refers to the outgoing null geodesics. Now, expanding
 the above equation around the horizon, we obtain
\begin{eqnarray}
\frac{dR}{dt}&=&\sqrt{1+f}-\sqrt{1+f}\sqrt{1-\frac{R-2m}{R(1+f)}}
\nonumber
\\
&\cong&\frac{R-2m}{2R\sqrt{1+f}},
\end{eqnarray}
where $R - 2m$ is assumed to be a non-zero small quantity.  Using the results
for the radial null geodesics, we obtain for the imaginary part of the action
corresponding to the LTB metric
\begin{eqnarray}
\label{pw-sy} {\textrm{Im}} S=
{\textrm{Im}}\int_{R_{\textrm{in}}}^{R_{\textrm{out}}}p_R dR
\nonumber\\
={\textrm{Im}}\int_{R_{\textrm{in}}}^{R_{\textrm{out}}}\int_0^H\frac{-dH'}{\dot
R} dR
\nonumber\\={\textrm{Im}}\int_{R_{\textrm{in}}}^{R_{\textrm{out}}}
\int_0^H\frac{2R\sqrt{1+f}(-dH')}{ R-2m} dR,
\end{eqnarray}
where we have used the above expansion for $\frac{dR}{dt}$ up to the
first order of $R - R_H = R - 2m$.  The corresponding Kodama
invariant energy of the particle $\omega=K^iS_i=\sqrt{1+f}\partial_t
S$ is then calculated to be $\omega=(\sqrt{1+f}) dH'$. We then use
the expansion of $R-2m(r)$ in the form
\begin{eqnarray}
\label{sy-ex} R-2m(r)&=&\nonumber\\
(1-2\frac{dm}{dR})(R-R|_H)-2\frac{dm(r)}{dt}(t-t|_H)&=&\nonumber\\((1-2\frac{dm}{dR})-\frac{2}{\frac{dR}{dt}|_{null}}\frac{dm(r)}{dt})(R-R|_H).
\end{eqnarray}
Changing the variables $(t,R)$ to $(t,r)$, makes it easier to use
the expression for the surface gravity in synchronous coordinates.
Putting
 $\frac{dm}{dR}=\frac{m'}{R'}$, $\frac{dm(r)}{dt}=m'\frac{\partial
r}{\partial t}|_{R=cont}$, $\frac{\partial r}{\partial
t}|_{R=cont}=-\frac{\dot{R}}{R'}$, and
$\frac{dR}{dt}=2\dot{R}$ in (\ref{sy-ex}) and (\ref{pw-sy}), we
obtain for the imaginary part of the action
\begin{eqnarray}
\label{pw-phys} Im S=\frac{w
\pi}{(\frac{1}{2R}-\frac{m'}{2RR'})}=\frac{w \pi}{\kappa_H},
\end{eqnarray}
 leading to the Hawking temperature (\ref{sgrav}):
\begin{equation}
T=\frac{\hbar \kappa_H}{2\pi}.
\end{equation}
This Hawking radiation temperature is the same as the one calculated in the previous section using Hamilton-Jacobi approach
\cite{hay0906}. The definition of the surface gravity used here has been essential
to arrive at this result indicating it to be more useful than the other definitions
in the literature \cite{nielsen}.\\
Note that according to (\ref{ltbe00}) and (\ref{ltbph}), the term
$R-2m$ and accordingly the metric factor of $dt^2$, i.e.
$\frac{\dot{R}^{2}}{1+f}-1$, vanishes on the horizon, leading to
$\frac{dR}{dt}=2\dot{R}$. A fact which has not to be assumed while
calculating the radial null geodesics, as has been done in \cite{hay0906}.
Otherwise it would result in the vanishing of the imaginary part of the action and therefore no
tunneling.

\section{First Law of the  Dynamical LTB  Black Holes}

 Using the tunneling approach for the Hawking radiation, we formulate now a first law of the LTB black hole. Consider
the following invariant quantity in the
 normal space (\ref{vanzo}): 
 \beq
 T^{(2)}=\gamma^{ij}T_{ij}\,,
 \eeq
where $T_{ij}$ is the normal part of energy momentum tensor. Now, using the invariant surface gravity
term in LTB given by (\ref{sgrav}), it is easy to show that on the
dynamical horizon of our LTB black hole we have \\
 \beq
 \kappa_H=\frac{1}{2R_H}+8\pi R_H T^{(2)}_H\,,
 \label{vanzo}
 \eeq
where we have used $T_{00}=-\rho$ according to the section (3.1) and Einstein equations (\ref{ltbe00}). The horizon area, the
areal volume associated with the horizon, and their respective
 differentials are then given by
  \beq \mathcal A_H = 4\pi R_H^2\,,\qquad d \mathcal
 A_H=8\pi R_H
 dR_H\,, \eeq \beq V_H=\frac{4}{3}\pi R_H^3\,,\qquad d V_H=4\pi R_H^2 dR_H\,.
 \eeq
 Substitution from above leads to

 \beq \label{se-l} \frac{\kappa_H}{8 \pi}d \mathcal A_H
 =d\left(\frac{R_H}{2}\right) + T_H^{(2)} dV_H\,. \eeq
Introducing the Misner-Sharp energy at the horizon, i.e. $ m = R_H/2$,  this can be recast in the
form of a first law:
 \beq
 dm=\frac{\kappa_H}{2 \pi}
d\left(\frac{\mathcal A_H}{4}\right) -T_H^{(2)} dV_H =
\frac{\kappa_H}{2 \pi} ds_H -T_H^{(2)} dV_H\,,
 \eeq
where $s_H=\mathcal A_H/4\hbar$ generalizes
 the Bekenstein-Hawking black hole entropy.\\

To see how this black hole first law is related to the Hawking
radiation, we concentrate on two conserved currents which can be
introduced in for LTB black hole. The first one is due
to the Kodama vector $K^a$, and the corresponding conserved charge
given by the area volume $V = \int_\sigma K^a d\sigma_a = 4 \pi
R^3/3$, where $d\sigma_a$ is the volume form times a future directed
unit normal vector of the space-like hypersurface $\sigma_a$. The
second one may be defined by the energy-momentum density $j^a=T_b^a
K^b$ along the Kodama vector, and its corresponding conserved charge
$E =-\int_\sigma j^a d\sigma_a$ being equal to the Misner-Sharp
energy. The total energy inside the apparent horizon can then be
written as $E_H = R|_H/2$, which is the Misner-Sharp energy at the
radius $R=R|_H$ of our LTB black hole. The energy outside the
region can be expressed as $E_{>H} =- \int_\sigma T_b^a K^b
d\sigma_a $, where the integration extends from the apparent horizon
to infinity. We may therefore express the total energy of the
spacetime as
 \beq
 E_t=R|_H/2-\int_\sigma T_b^a K^b d\sigma_a.
\eeq Consider now a tunneling process. The initial state before the
tunneling defined by $R = R|_H$ having an energy $E_t^i$, and the
final one after the tunneling defined by $R = R|_H+\delta R|_H$
having the energy $E_t^f$. According to the energy conservation, the
Kodama energy change between the final and initial states of the
tunneling process is then calculated to be \beq w = dE_t =
E_t^f-E_t^i=\frac{\delta R|_H}{2}-\rho dV.  \eeq

Substituting from (\ref{pw-phys}) and (\ref{se-l}) in the above
equation, we obtain the tunneling rate $\Gamma \sim e^{\frac{-2 Im
S}{\hbar}} = e^{\frac{-1}{4\hbar}\int_{s_{i}}^{s_f} d\mathcal
A_H}=e^{\Delta s}$, with $\Delta s=s_f - s_i$ being the entropy
change. Our discussion shows that the tunneling rate arises as a
natural consequence of the unified first law of thermodynamics $dE_H
= Tds - T_H^{(2)} dV_H$ at the apparent horizon.

\section{Non-thermal radiation from the LTB black hole}

The question of how the formulas for black hole radiation are
modified due to the self-gravitation of the radiation is dealt with
by Kraus and Wilczek in \cite{kraus95}. There it is shown that the
Hawking radiation is non-thermal when the energy conservation is
enforced. The particle in the particle hole system is treated as a
spherical shell to have a workable model with the least degrees of
freedom. The radiating black hole of mass $M$ will be modeled as a shell
with the energy $\omega$  around the hole having the energy $M(r)-\omega$. We
are going to adapt this model to our LTB black hole. Note first that due to
the fact that our LTB black hole model is asymptotically Friedman-like, the
ADM mass is not available to be fixed there. We therefore turn to the
quasi-local Misner-Sharp mass. Let us then calculate the tunneling amplitude
using this modification:

\begin{eqnarray}
\label{pw-sy-back}
{\textrm{Im}} S&=&
{\textrm{Im}}\int_{R_{\textrm{in}}}^{R_{\textrm{out}}}p_R dR
\nonumber
\\
&=&{\textrm{Im}}\int_{R_{\textrm{in}}}^{R_{\textrm{out}}}\int_0^H\frac{-dH'}{\dot
R} dR \nonumber
\\
&=&{\textrm{Im}}\int_{2M(r)}^{2(M(r)-\omega)} dR \int_0^H\frac{-dH'}{\dot
R}.
\end{eqnarray}
Carrying out the first integral, we obtain
\begin{eqnarray}
{\textrm{Im}} S=\int_0^H \frac{\pi dH'
}{(\frac{1}{4M-4H'}-\frac{m'}{(4M-4H')R'})}\simeq \nonumber
\\
\frac{\pi H(1-\frac{H}{2M})
}{(\frac{1}{4M}-\frac{m'}{4MR'})}=\frac{\pi w(1-\frac{\omega}{2M})
}{\kappa_H},
\end{eqnarray}
where we have used the fact $\frac{\omega}{2M}<<1$ in the last step to
expand the integrand and carry out the integral. The result shows the
non-thermal character of the radiation.\\
It has been shown in \cite{man-ash} that the boundary of the LTB
black hole becomes a slowly evolving horizon for  $R'>>1$, with the
surface gravity being equal to $\kappa_H=\frac{1}{4M}$, and
an infinite redshift for the light coming out of this horizon. Therefore,
using the above equation, the tunneling probability has the same
form as in the case of Schwarzschild, i.e. $\Gamma\sim exp[-8 \pi
\omega (M-\frac{\omega}{2}) ]$. Having this form of the tunneling
probability we may refer to Zhang et al. \cite{cai-inf09} who have
shown that this form of the non-thermal radiation leads to the
conservation of the total entropy.

\section{Conclusions}

Within a research program to understand more in detail the LTB cosmological black hole, its
similarities and differences to the Schwarztschild black hole, we have calculated the
Hawking radiation from this dynamical black hole by using the tunneling methods suitable for dynamical cases. It turns
out that for the LTB black hole the Hamilton-Jacobi and the radial null geodesic-method both lead to the same tunneling rate.
It turns out that it is not the event horizon but the future outer tapping horizon that contributes to this Hawking
radiation.\\
Formulation of a first law for the LTB black hole and the tunneling amplitude
show that the tunneling rate has a direct relation to the change of the LTB black hole
entropy. Assuming the energy conservation for the LTB black hole's slowly
evolving horizon, we show that the radiation is non-thermal and that the
entropy is conserved during the radiation.\\

\end{document}